# Weather instabilities as a warning sign for a nearby climatic tipping point?

François Louchet
Chemin du Pré Roudon, 384210 St Martin d'Uriage (France)
(formerly at "Laboratoire de Glaciologie et de Géophysique de l'Environnement", Grenoble)
francois.louchet@yahoo.fr
https://sites.google.com/site/flouchet/

**Abstract**

Using in a simple way the theory of non linear dynamical systems, we show that increasing climatic instabilities may be a qualitative warning sign for the occurrence of a nearby bifurcation, yielding a discontinuous and sudden climate tipping towards an unknown and unpredictable state. The possibility of an accurate prediction of the occurrence time of such a transition is also discussed in terms of the approach of a critical point.

**Introduction**

Present predictions of climatic evolution are based on models involving numerous parameters and physical laws, and on known climatic situations recorded in past and present times, as recently detailed in [1]. Unfortunately, most involved mechanisms are poorly known or ignored, and data are also sometimes biased, due to different measurement techniques by different organizations, various averaging or filtering procedures, etc... In addition, current predictions are usually based on continuous extrapolations of recent evolutions, using different (i.e. more or less optimistic or pessimistic) scenarii [1]. Furthermore, by contrast with past climatic evolutions, recent global warming is characterized by short time forcing, that may yield rapid and unexpected reactions of the coupled atmospheric-oceanic system.

On the other hand, the possible occurrence of abrupt climatic transitions, usually named "tipping points" in climatology, as a response to continuous evolutions of driving parameters, was recently discussed in the literature [2-6], and mentioned in the 2013 IPCC report [1]. Such transitions are reminiscent of critical phenomena occurring in complex systems, defined as ensembles of a large number of interacting entities, in which the global behaviour may significantly differ from that of isolated elements.

In all cases, system tipping results from departure from a stable attractor. According to Ashwin et al. [2], tipping points in open systems could be classified in three different types, resp. labelled B (for "bifurcation"), N (for "noise") and R (for "rate dependent"). Actual tipping events may be described in some cases by combinations of several of the above types. However, the possible observation of fluctuations at the vicinity of a tipping point are strongly reminiscent of a B-type, possibly associated with a N-component, as detailed in the next section in the case of climate tipping.

The present paper aims at giving in a simple way a qualitative hint on the possible occurrence of an abrupt and discontinuous climatic change in the near future, driven by continuous evolution of "control" parameters. The possibility of an accurate prediction of the occurrence time of such a tipping point is also discussed in terms of climatic instabilities that may announce the approach of such a transition.

**Fluctuations in the vicinity of a critical point**

Since we are interested here in fluctuations as warning signs for climatic tipping approach, we shall focus the present analysis on B-type tipping points, i.e. bifurcations, that exhibit such precursory signs.

Let us consider a complex system whose state is described by a function $Z(C,t)$ of time $t$, parametrized by a set of parameters globally represented by $C$. In the theory of critical phenomena, $Z$ is called "order parameter", and $C$ "control parameters".
We are interested in the evolution of $Z$ as a function of time, for different values of $C$.

For instance, if we are interested in the "simple" case of time evolution of a crack in a material, $Z$ may be the crack size, and $C$ the applied stress. Increasing $C$ increases $Z$ continuously, until a critical point is reached, where the crack becomes suddenly unstable and consequently uncontrolled (i.e. $Z$ diverges), leading to material failure.

Another example could be the evolution of the Gulf Stream "position" $Z$ off european shorelines, with $C$ including the average $CO_2$ concentration in the atmosphere, that of $CH_4$, the average atmosphere temperature, etc... In a third example, discussed hereafter, $Z$ should be the average atmosphere temperature, and $C$ the green house gases concentration.

In the general case, the time variations of $Z(C,t)$ can be studied in a very simple way, representing its time derivative $dZ/dt$ as a function of $Z$ itself. In a non-linear dynamic system, $Z$ is usually a fairly complicated (though in principle continuous) function, and $dZ/dt$ may intersect the $Z$ axis at several points, as shown schematically in fig. 1 (bottom blue curve). Since such points correspond to zero values of the derivative $dZ/dt$, they are equilibrium points, called "fixed points". We are now interested in the system behaviour at the vicinity of such fixed points.

Fixed points at which the curve slope is negative ($A_1$ and $A_2$ in fig. 1) correspond to stable states of the system, and are called "attractors". Indeed, if the system starts from a non-equilibrium state, for instance from a $Z$ value slightly larger than that at $A_1$, the corresponding value of *dZ/dt* is negative, which drives the system towards $A_1$, and conversely if the starting *Z* value was lower than that at $A_1$.

In a same way, fixed points for which the curve slope is positive ($R_1$ on fig. 1) correspond to unstable states of the system, and are called "repulsors".

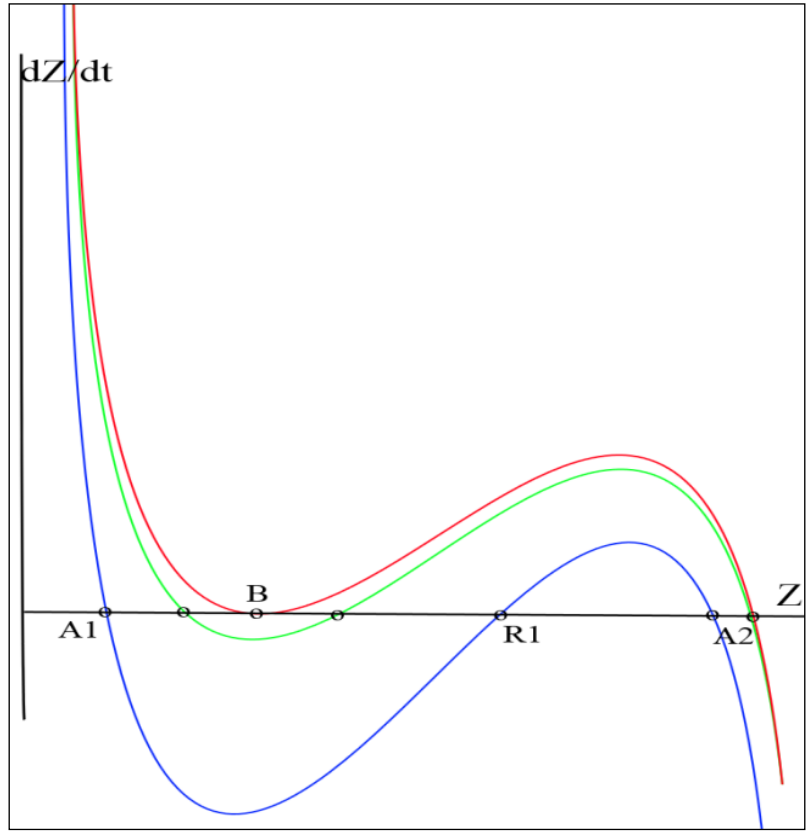

Fig. 1: Schematic example of the evolution of a complex system. $A_1$ and $A_2$ are attractors, $R_1$ is a repulsor. A bifurcation occurs when $A_1$ and $R_1$ merge at B, that results in a jump of the system to the next attractor $A_2$. (see text).

On the blue curve of fig.1, the zone between $-\infty$ and $R_1$ is the attraction basin of $A_1$, and that between $R_1$ and $+\infty$ is the attraction basin of $A_2$. These attraction basins are obviously separated by the repulsor $R_1$.

Let us now consider a system in a stable state like $A_1$. If control parameters *C* gradually change, the curve in fig. 1 would be modified. Let us suppose for instance that the curve deforms and (or) shifts upwards. The attractor $A_1$ and the repulsor $R_1$ become closer together (intermediate green curve), and eventually merge (top red curve) for a critical set of control parameters called *C+* corresponding to a critical value $Z_c$ of *Z*. Beyond *C+*, the attraction basin of $A_1$ would disappear, and the system would suddenly jump to higher *Z* values, up to the nearest attractor $A_2$. Such a behaviour shows that a continuous change of control parameters may in some cases yield a discontinous change in the system state. It is a typical example of a bifurcation [6].

We shall now investigate more closely typical phenomena that may occur at the vicinity of such a bifurcation B. If a "noise" is introduced, due to fluctuations of minor and uncontrolled parameters that may not have been taken into account in *C*, the system would be randomly drawn aside its theoretical equilibrium state $A_1$, and brought back towards $A_1$ due to the attractive character of $A_1$.

Now, as $A_1$ and $R_1$ are brought nearer to one another, the curve slopes *d/dZ(dZ/dt)* at $A_1$ and $R_1$ decrease in absolute values (intermediate green curve) and vanish at merging (top red curve). As a consequence, the driving force bringing the system back towards equilibrium would decrease with the absolute value of the derivative *d/dZ(dZ/dt)* at the vicinity of the attractor, and would vanish at the bifurcation point B. As a consequence, the response of the system becomes more sluggish, which is known as the "critical slowing down" [7]. For the same reason, the same noise would yield increasing *Z* fluctuations as the bifurcation becomes closer. A discrete jump of *Z* would occur when Z fluctuations tend to infinity, i.e. at the bifurcation.

Another characteristic of such a sudden transition for a critical set of parameters *C+* is that it is hardly reversible. In the present simple example indeed, starting from $A_2$, if the evolution of control parameters *C* is reversed, the curve would shift downwards, but the system would remain locked on $A_2$ even if *C* recovers its transition value *C+*. In order to bring the system back to the initial attractor $A_1$, it would be necessary to shift the curve significantly further down, to the position where $A_2$ and $R_1$ merge in turn, for a value *C=C-* of control parameters. The system would then experience another transition, leaving the attraction basin of $A_2$ that would have disappeared, falling into the attraction basin of $A_1$, thus achieving an hysteresis loop. This type of irreversibility is not a question of inertia, as sometimes asserted, but results from the fact that the conditions for tipping back to the initial attractor usually differ significantly from those that allowed the initial forward jump.

The conclusions are threefold:
i) the approach of the bifurcation is characterized by fluctuations of increasing amplitudes
ii) such amplitudes diverge at the bifurcation, allowing the system to escape in the direction of the previous continuous drift, towards the nearest stable equilibrium state.
iii) this transition is hardly reversible, and characterized by an hysteretic behaviour.

**Possible application to climatic changes**

The properties of non-linear dynamical (complex) systems discussed above are quite general. A possible application of the present analysis is the case where the function *Z* is the average atmosphere temperature. For the sake of simplicity, the large variety of control parameters are supposed here to be constant, except for the atmospheric average $CO_2$ concentration.

We start from a stable system, staying in an attractor like $A_1$. When $CO_2$ concentration increases, so does temperature. This is an observational fact, ascribed to the well known green house effect. It means that the attractor $A_1$ in fig. 1 continuously drifts to the right.
The observation of increasing amplitude fluctuations for temperatures (extreme hot or cold periods) or for more or less directly connected events (e.g. storms) may suggest that the local shape of the curve and its time evolution are similar to that of fig. 1 (a,b,c), i.e. that its local slope at the attractor decreases in absolute values. The system is therefore probably approaching a bifurcation, that may suddenly propel it towards a new attractor, with a significantly different temperature. Such a discrete jump would go in the same direction as the continuous trend preceding the transition, i.e. towards higher temperatures. But, even if the occurrence of such a transition may be predicted, the new equilibrium temperature might be significantly larger than what any continuous extrapolation of past climates could forecast, and would be totally unpredictable.

It is therefore clear that continuous extrapolations of past climatic changes in order to predict future evolutions are all the more untrustable as fluctuations of increasing amplitudes are observed. In this case, the system is likely to meet sooner or later a bifurcation yielding a discontinuous jump towards a new equilibrium, whose characteristics are difficult (or impossible) to describe precisely. Such a jump would be irreversible, in the sense that even bringing control parameters ($CO_2$ concentration in the present simplified example) back to values they had before the transition would not be sufficient to bring the system back to the initial equilibrium.

It is worth mentioning that such an abrupt transition, named Paleocene-Eocene Thermal Maximum (PETM), occurred 56 Myrs ago. Unlike many previous climate changes due to Earth orbit variations, PETM resulted from a considerable increase of green house gases essentially due to volcanic emissions [8]. It was enhanced by numerous positive feedbacks, as for instance outgassing of methane (20 to 25 times more efficient than $CO_2$ in terms of greenhouse effect) from methane calthrates into the atmosphere. In that sense, PETM is fairly similar to the present warming phenomenon. By contrast, the peak rate of carbon-based gases (i.e. $CO_2$ and $CH_4$) addition to the atmosphere-ocean system during PETM, expressed in terms of carbon mass, was estimated in the range 0.3-1.7 Pg C/yr (0.3-1.7 $10^{15}$ g C/yr), i.e. about 15 times less than the present carbon emission rate [9]. It nevertheless yielded a temperature increase estimated between 2 and 9$^o$C (fig. 4 in [9]). Despite the fact that the conditions may be somewhat different between PETM and present warming events, this estimate may give an order of magnitude of what could happen if a tipping event had to take place.

In order to check whether such a transition is likely to occur in a near future and predict when it would take place, the ideal solution should be to build a complete physical model, with initial conditions chosen at any moment in the past, but with an infinite accuracy. It is well known however that, due to the very large sensitivity to initial conditions of complex systems, the faintest inacurracy in initial data would totally invalidate the predictions, as initially shown by Poincaré in 1890.

Another solution should be to take the problem the other way round. As mentioned above, the increasing instabilities observed in the system characterize the proximity of a bifurcation-type tipping point. Such phenomena at the vicinity of bifurcations have been extensively studied in the case of geological failures, as earthquakes [10], volcanic eruptions [11], rockfalls [12], landslides [13], or cold glacier ruptures [14, 15].

The approach of the critical point usually obeys a power law, characteristic of scale invariant phenomena [1]. In the case of cold glacier ruptures for instance, it can be written:

$$V(t) = V_o + a(t_c - t)^p \qquad (1)$$

where $V(t)$ is the glacier surface velocity at time $t$, $V_o$ a constant, $t_c$ the failure time, and $p$ (<0) the power law exponent. The glacier velocity $V(t)$ diverges at $t=t_c$. Fitting parameters of eq. (1) allows a prediction of the failure time $t_c$ [14].

However, owing to the improved accuracy of measurements, it was shown recently that glacier displacements can be alternatively fitted by a fairly similar law, where the power law term is modulated by a log-periodic function [14-17]. The time variations of the displacements $s(t)$ at the vicinity of the bifurcation can be written in this case [15]:

$$s(t) = s_0 + u_s t - a(t_c - t)^m \left[ 1 + C \sin\left( 2\pi \frac{\ln(t_c - t)}{\ln \lambda} + D \right) \right] \qquad (2)$$

where $s_0$, $a$, $C$, $D$ and $\lambda$ are constants, $u_s t$ is a constant velocity drift, and $a(t_c-t)^m$ is a power-law acceleration with a power law exponent $m=p+1$ (<1), modulated by a log-periodic oscillatory term. This equation actually describes oscillations with increasing amplitudes and frequency as the system approaches the critical point at $t=t_c$. An interesting feature is the frequency divergence at $t=t_c$, due to the logarithmic term. Fitting this equation on field measurements, and more particularly on observed oscillations, allows a prediction of the time $t_c$ at which the transition will occur, with an improved accuracy as compared to eq. (1). Despite the fact that the origin of such an oscillatory behaviour is still debated (p. 209 in [15], this procedure yields cold glaciers breakdown predictions with an accuracy of a few days [14,15].

Transposition of eq. (2) to climate is not straightforward, since log-periodic oscillations are usually recognized to characterize the so-called "Discrete Scale Invariance" (DSI) [1] ([17] paragr. 5.3.2, [18]), that can be described using complex power law exponents. It should first be checked whether climatic fluctuations are of a similar nature as such oscillations, which is not obvious, but not impossible. It seems indeed that DSI is fairly frequent in various domains ([17], paragr. 5.4,. 6.2), and more particularly as the system departs from an attractor ([17], paragr. 6.5). If application of eq. (2) to climate proved to fail, a less accurate tipping point forecast could be tempted using eq. (1).

**Summary and conclusion**

Most present predictions of climate evolution are based on continuous extrapolations of past observed trends. As a consequence, measures that are and will be taken in order to slow down the present climatic evolution might be undersized and unefficient.

On the basis of the theory of non-linear dynamical systems, it is clear that observed increasing climatic instabilities may be a qualitative warning sign for the occurrence of a nearby bifurcation, yielding a discontinuous and sudden climate tipping. Such a mechanism would bring the system, through an irreversible jump, towards an unknown and unpredictable state, quite different from what any continuous extrapolation of present data may predict.

If this was to be the case, a forecast of the occurrence time of the transition would be of interest. The atmosphere being an out-of-equilibrium complex system, the transposition to climate evolution of methods used to describe the approach of critical points, particularly in geophysics, would be in principle applicable.

Unfortunately, presently available climatic data do not seem to be sufficient for an accurate prediction of both the occurrence and time of such a possible transition. There is an urgent need for precise and reliable data that could allow such models to be implemented.

**Notes**

(1) The concept of scale invariance is fundamental in the study of critical phenomena. A so-called "scale invariant" function $y(x)$ is such that $y(x)/y(\lambda x)$ keeps a constant value $\mu$ when $x$ is multiplied by any factor $\lambda$:

$$\frac{y(\lambda x)}{y(x)} = \frac{(\lambda x)^m}{x^m} = \lambda^m \text{ with } \lambda^m = 1/\mu$$

For a given value of the exponent $m$ (that characterizes the studied physical phenomenon), the value of $y$ at two different scales only depends on the the ratio $\lambda$ between both scales.
The Discrete Scale Invariance is a less general scale invariance, in that it is valid only for an infinite but denombrable discrete ensemble of $\lambda$ values. This is the case for instance for the "Cantor set".
In the case of cold glaciers, it seems intuitive that the DSI should be related to the discrete fractal nature of the crack network that leads to the final failure, that introduces a privileged "duplication period" somewhat similar to the Cantor set [ https://en.wikipedia.org/wiki/Cantor_set ].

**Acknowledgements**

The author is grateful to his colleague Dr Gerhard Krinner, IPCC member, and to Dr Jérome Faillettaz, Zurich University, for fruitful discussions.
**References**

[1] 2013 IPCC report.

[2] Ashwin P., Wiekzorek S. Vitolo R., Cox P., Phil. Trans. R. Soc. A (2012) **370**, 1166-1184, doi: 10.1098/rsta.2011.0306.

[3] Abrupt Climate Changes. Inevitable Surprises, The National Academies Press (2002) ISBN: 978-0-309-07434-6, DOI: 10.17226/10136.

[4] Lenton T.M., Livina V.N., Dakos V., van Nes E.H., Scheffer M., Early warning of climate tipping points from critical slowing down: comparing methods to improve robustness. Phil. Trans. R. Soc. (2012), A 370, 1185-1204, doi:10.1098/rsta.2011.0304.

[5] Lenton T.M., Early warning of climate tipping points, Nature Climate Change (2011) 1, 201–209 doi:10.1038/nclimate1143, Published online, 19 June 2011 .

[6] Lenton T.M., Livina V.N., Dakos V., Scheffer M., Climate Bifurcations during the Last Deglaciation? Clim. Past (2012), 8, 1127-1139.

[7] http://www.early-warning-signals.org/theory/what-is-a-critical-transition/

[8] https://en.wikipedia.org/wiki/Paleocene%E2%80%93Eocene_Thermal_Maximum

[9] Cui Y., Kump L.R., Ridgwell A.J., Charles A.J., Junium C.K., Diefendorf A.F., Freeman K.H., Urban N.M. & Harding I.C. (2011). "Slow release of fossil carbon during the Palaeocene–Eocene Thermal Maximum". Nature Geoscience. **4** (7): 481–485. *doi:10.1038/ngeo1179*].

[10] Bufe, C. G., and D. J. Varnes, Predictive modeling of the seismic cycle of the greater San Francisco Bay region, J. Geophys. Res. (1993), 98 (B6), 9871–9883, doi:10.1029/93JB00357],

[11] Voight, B., A method for prediction of volcanic eruptions, Nature(1988), 332(6160), 125–130.

[12] Amitrano, D., J. R. Grasso, and Senfaute G., Seismic precursory patterns before a cliff collapse and critical point phenomena, Geophys. Res. Lett. (2005), 32, L08314, doi:10.1029/2004GL022270.

[13] Sornette, D., Helmstetter A., Andersen J., Gluzman S., Grasso J.-R., and Pisarenko V., Towards landslide predictions: Two case studies, Physica A (2004), 338 (3-4), 605–632, doi:10.1016/j.physa.2004.02.065

[14] Flotron, A., Movement studies on hanging glaciers in relation with an ice avalanche, J. Glaciol. (1977), 19 (81), 671–672.

[15] Faillettaz J., Funk M. and Vincent C., Avalanching glacier instabilities: Review on processes and early warning perspectives, Rev. Geophys. (2015), 53, 203–224, doi:10.1002/2014RG000466, open access at: http://onlinelibrary.wiley.com/doi/10.1002/2014RG000466/pdf

[16] Pralong, A., Oscillations in critical shearing, application to fractures in glaciers, Nonlinear Processes Geophys. (2006), 13, 681–693.

[17] Sornette D., Discrete scale invariance and complex dimensions, Physics Reports (1998), 297, 239-270.

[18] Sornette D. and Sammis C. G., Complex critical exponents from renormalization theory of earthquakes: Implications for earthquake predictions, J. Phys. I Fr. (1995),5 (5), 607–619, doi:10.1051/jp1:1995154.